# Crowdsourced science: sociotechnical epistemology in the e-research paradigm

**David Watson**[1] · **Luciano Floridi**[2]

Received: 30 April 2016 / Accepted: 29 September 2016

**Abstract** Recent years have seen a surge in online collaboration between experts and amateurs on scientific research. In this article, we analyse the epistemological implications of these crowdsourced projects, with a focus on Zooniverse, the world's largest citizen science web portal. We use quantitative methods to evaluate the platform's success in producing large volumes of observation statements and high impact scientific discoveries relative to more conventional means of data processing. Through empirical evidence, Bayesian reasoning, and conceptual analysis, we show how information and communication technologies enhance the *reliability*, *scalability*, and *connectivity* of crowdsourced e-research, giving online citizen science projects powerful epistemic advantages over more traditional modes of scientific investigation. These results highlight the essential role played by technologically mediated social interaction in contemporary knowledge production. We conclude by calling for an explicitly sociotechnical turn in the philosophy of science that combines insights from statistics and logic to analyse the latest developments in scientific research.

✉ David Watson
d.watson@qmul.ac.uk

1 Queen Mary University of London, London, UK

2 Oxford Internet Institute, University of Oxford, Oxford, UK

## 1 Introduction

Experts and amateurs have been collaborating on so-called 'citizen science' projects for more than a century (Silvertown 2009). Traditionally, such projects relied upon volunteers to participate in data *collection*. In more recent years, the spread of information and communication technologies (ICTs) has allowed users to become increasingly involved in data *analysis*. Early online citizen science initiatives made use of participants' spare processing power to create distributed computing networks to run simulations or perform other complex functions (Anderson et al. 2002). The latest wave of citizen science projects has replaced this passive software approach with interactive web platforms designed to maximise user engagement. Utilising fairly simple tools provided by well-designed websites, amateurs have helped model complex protein structures (Khatib et al. 2011a, b), map the neural circuitry of the mammalian retina (Kim et al. 2014), and discover new astronomical objects (Lintott et al. 2009; Cardamone et al. 2009). As of December 2015, citizen science project aggregator SciStarter links to over a thousand active projects (SciStarter 2015).

What are the philosophical implications of this new brand of crowdsourced e-research? Sociologists have studied the demographics and motivations of virtual citizen scientists for years (e.g., Nov et al. 2011; Rotman et al. 2012; Raddick et al. 2013), while data scientists have extensively examined the mechanics of user contributions to such sites (e.g., Kawrykow et al. 2012; Ponciano et al. 2014; Franzoni and Sauermann 2014). Philosophers, however, have so far been silent on these methodological developments. In this article, we argue that a close examination of crowdsourced e-research reveals important lessons for epistemology and philosophy of science.

Virtual citizen science labs constitute large sociotechnical systems in which professionals, volunteers, and digital technologies come together to pursue three important epistemic goals:

(1) *Reliability* The designers of citizen science websites employ numerous quality control measures to ensure that user contributions are accurate and precise.
(2) *Scalability* Hundreds of thousands of volunteers from around the world regularly participate in citizen science projects, analysing unprecedented volumes of data for a wide variety of scientific studies.
(3) *Connectivity* Information and communication networks unlock the distributed knowledge of large epistemic communities by establishing numerous channels that allow users to confer with one another and direct information toward one or several central nodes.

In what follows, we present empirical evidence that crowdsourced e-research is uniquely reliable, scalable, and connective. We argue that these properties are essential for the promotion of scientific knowledge, and therefore that any system that maximises all three not only constitutes a major methodological advancement, but merits close philosophical attention. We conclude that the success of virtual citizen science underscores the irreducibly sociotechnical nature of all scientific inquiry.

Following an overview of this paper's methods in Sect. 2, we proceed to examine the structural mechanics of contemporary citizen science in Sects. 3–5, with an emphasis on the epistemic advantages afforded by web-enabled mass collaboration. Our findings indicate that such projects tend to generate more observations and higher quality discoveries than similar studies using traditional methods. The significance of these results goes far beyond the limits of citizen science. We close in Sect. 6 with a review of our findings and a proposal for further research in sociotechnical epistemology.

## 2 Motivation and methods

Suppose Albert and Niels are rational agents with opposing views on which of two mutually incompatible scientific hypotheses is correct. Let us assume that fundamental disagreements between the two men are negligible—they play by roughly the same epistemic rules and are each willing to concede a point in the face of sufficient evidence or compelling arguments. Yet despite their concordance on basic principles, they just cannot seem to agree on this particular case. What might explain this (common) scenario?

Say Niels happens to be right in this instance. Then at least one of three possibilities accounts for his success: (a) he got lucky; (b) he had better evidence; or (c) he had a better understanding of the evidence. If our goal is to find the most fruitful strategies for scientific inquiry, then explanation (a) is irrelevant. Options (b) and (c) are more interesting, however. The first highlights the importance of good evidence, which can be split into data quality and quantity (Floridi and Illari 2014). The second suggests that even in the face of identical evidence, superior results are achieved by the agent who does a better job of finding the underlying structure behind a given set of observations.

In one of the seminal works of social epistemology, Goldman (2003) sets out to evaluate various systems for making and improving judgments through different forms of testimony. Central to his project is the notion of 'veritistic value', a measure of one's degree of knowledge or truth possession with respect to a proposition. Let $T$ stand for the truth-value function such that $T(p) = 1$ iff $p$ is true and $T(p) = 0$ iff $p$ is false. Let $C$ stand for agent $A$'s credence function such that $C_A(p) = 1$ iff $A$ is certain that $p$ and $C_A(p) = 0$ iff $A$ is certain that $\sim p$. Then the veritistic value of $A$'s judgment with respect to $p$ may be defined as a function $V$ such that $V_A(p) = 1 - |T(p) - C_A(p)|$.[1] In our motivating example above, Niels' judgment was of higher veritistic value than Albert's.

We submit that for a wide array of projects throughout the natural sciences, crowdsourcing offers the best available method for maximising the expected veritistic value of researchers' hypotheses. Thoughtful web protocols and global Internet access ensure high data quality and quantity, while the sociotechnical network's topology pushes anomalous observations to the fore, thereby challenging experts to find the latent structure underlying the natural phenomena they study. This conclusion is derived from a combination of empirical findings and logical reasoning presented below. For the former, we draw primarily on data from and about Zooniverse, the world's largest citizen science web portal. For the latter, we adopt a Bayesian confirmation theoretic framework that borrows from the social epistemology of Goldman (2003) and the epistemic logic of Fagin et al. (1995).

[1] Goldman does not use these precise formulae, although they are implicit in his 'trichotomous scheme'. See Goldman (2003, Sect. 3.4, pp. 87–94).

Because a priori reflection alone is insufficient to substantiate our argument, we review Zooniverse's 2014 transaction logs and complete publication history to better understand the platform's internal mechanics and scientific output. With dozens of active projects and over 1.4 million subscribers worldwide, Zooniverse exemplifies the reliability, scalability, and connectivity of contemporary crowdsourced e-research we intend to analyse. The site began in July 2007 with a single project, Galaxy Zoo, which invited users (aka 'Zooites') to help classify the morphological properties of galaxies captured by the Sloan Digital Sky Survey (SDSS). Following the success of this inaugural venture, administrators (aka 'Zookeepers') expanded the site into a multi-project platform in December 2009. While the vast majority of Zooniverse projects are devoted to topics in the natural sciences, the site has recently branched out to include digital humanities initiatives as well. Figure 1 provides a breakdown of all 27 projects hosted by the platform in 2014.

Unlike the competitive games offered by designers of other popular crowdsourced science sites such as FoldIt (Cooper et al. 2010) and EyeWire (Kim et al. 2014), Zooniverse projects are based entirely upon *classifications*, be they of galaxies, whale calls, or ancient manuscripts. Each project starts with a simple set of instructions on how to classify the relevant digital artefacts, followed by a steady stream of raw data ready for processing (Simpson et al. 2014). As of December 2015, Zooniverse classifications have been the basis for 81 articles published in peer-reviewed journals, in addition to a handful of conference papers and book chapters (Zooniverse 2015).

We examined those publications for content and scientometric performance using Elsevier's Scopus database and the Thomson Reuters Institute for Scientific Information Journal Citation Reports. Web analytic data from Zooniverse's 2014 transaction logs were generously provided by the platform's administrators. Together, these sources provide the empirical basis for this paper's epistemological claims. Quantitative analysis was conducted in the R statistical environment (version 3.2.2), with significance levels for all tests uniformly fixed at $\alpha = 0.05$.

## 3 Reliability: the wisdom of the crowd

The success of any scientific study, crowdsourced or otherwise, crucially relies upon the reliability of its observations. How can we trust Zooniverse's classification data if they merely represent the uninformed opinion of a large community of amateurs?

### 3.1 Quality control protocols

The insight that groups are often better at producing knowledge than individuals is an old one. A formal proof of the claim was originally derived by Condorcet (1785), whose famous jury theorem states that given a defendant of uncertain guilt and a collection of jurors whose judgments are each better than random but less than perfect, the majority of jurors is always more likely to be correct than any individual juror. Moreover, the probability of a correct majority judgment approaches 1 as the jury size increases. An important corollary to Condorcet's jury theorem, however, is that opposite results will hold for worse than random jurors. That is, given a jury composed of individuals with a less than 0.5 chance of making accurate judgments, the majority is always more likely to be wrong than any individual juror, and the probability of a correct majority judgment approaches 0 as the jury size increases.

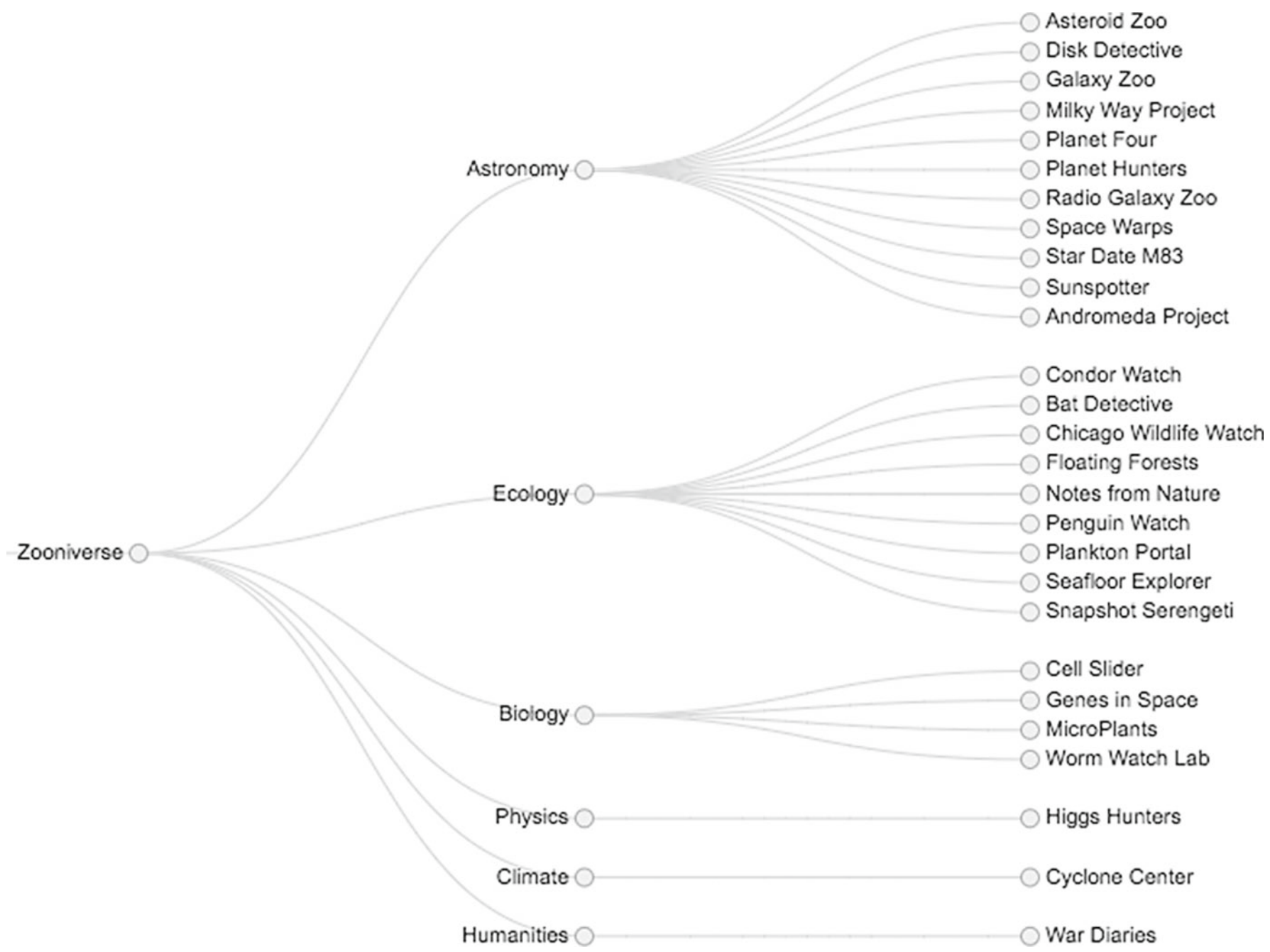

**Fig. 1** Dendrogram depicting the typological breakdown of all 27 Zooniverse projects active in 2014

The initial sceptical challenge to citizen science is motivated by something like the corollary to Condorcet's jury theorem (Collins 2014). Highly specialised subjects within the natural sciences are dominated by experts for good reason. Amateur views on particle physics or microbiology are probably wrong, these sceptics allege, and large groups of amateurs pooling their collective ignorance will surely do no better. The issue is perhaps best understood as a special case of the more general problem of testimony, upon which much of social epistemology turns.

With hundreds of thousands of users participating in any given crowdsourcing project, odds are that at least some will perform worse than random at certain data classification tasks. To stay on the right side of Condorcet's jury theorem, Zooniverse's administrators employ several strategies:

- *Design simplicity* Before a project is launched, Zookeepers ensure that tasks are simple and clearly explained to maximise potential contributors and minimise user error.
- *Automated filtering* Once a project is underway, algorithms filter classifications by user performance and community agreement across observations.

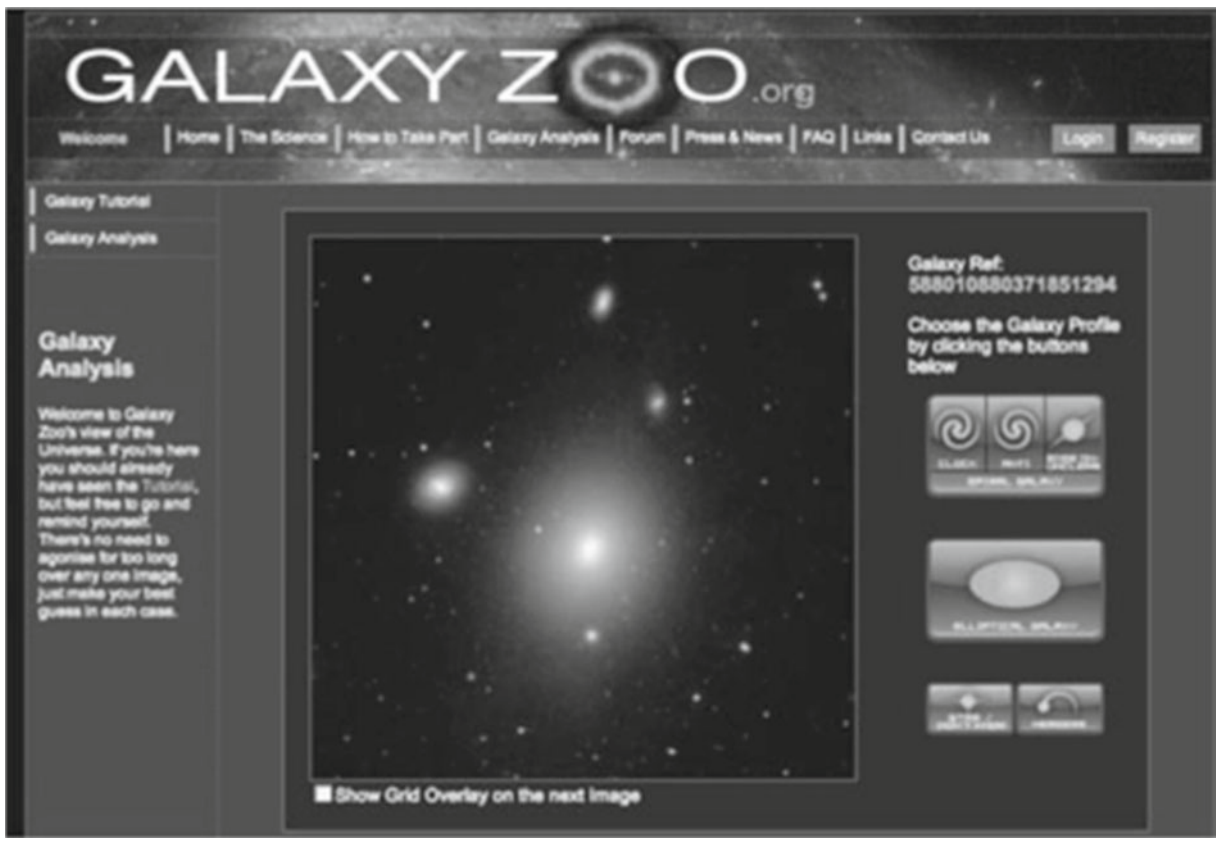

**Fig. 2** A screenshot from the original Galaxy Zoo website

- *Comprehensive review* Once a project is completed, classifications are weighted according to each user's tendency to be in the majority and full datasets are subject to expert review.

Together, these quality control measures have a profound impact on the scientific utility of amateur observations. To see how, consider the case of Galaxy Zoo.

Zooniverse's first project was a straightforward classification task, explained to new users in a brief tutorial that made no use of technical terminology. Volunteers were presented with paradigmatic examples of standard galaxy types and asked to determine to which type subsequent galaxies properly belonged. A total of six classifications were possible, with small schematic symbols of the available options permanently visible at the right of the screen (see Fig. 2). No experience in astrophysics was presumed, and in fact, with just a little practice, even young children could (and did) participate (Raddick et al. 2013).

Once users completed the tutorial, they were unknowingly subject to a probationary period during which they were presented with test data that Zookeepers considered unambiguous cases of their particular galactic morphologies. Classifications by those who failed to correctly identify 11 out of their first 15 images were not saved in the site's database (Lintott et al. 2008). This ensured that erroneous results from volunteers who misunderstood the instructions, experienced technical difficulties, or perhaps even maliciously sought to corrupt Galaxy Zoo's data, would not confound the project's findings.

As a further precaution, Zooniverse administrators designed a redundant website architecture in which numerous users reviewed each galaxy before it entered the project's catalogue. Objects were processed an average of 38 times each, allowing researchers to estimate the confidence of their conclusions by evaluating the extent of community consensus around particular classifications (Lintott et al. 2008).

Once the entire SDSS survey had been classified, Zookeepers applied a weighted voting schema in which each user's contributions were valued in proportion to the average popularity of her classifications.[2] A comparison of weighted and unweighted results revealed that, while there was practically no difference between the two scoring methods in terms of ultimate morphological selections, weighting user votes pushed tens of thousands of galaxies past researchers' 80 and 95 % consensus thresholds for entrance into 'clean' and 'superclean' morphological samples, respectively (Lintott et al. 2008).

A final, crucial step in Zooniverse's quality control protocol is the expert review of user observations. Examining Galaxy Zoo's results, researchers found significant over-classification of anti-clockwise spirality, probably due to the population's preference for right handedness (Land et al. 2008). Elliptical galaxies were also over-classified, most likely because spiral galaxies viewed at great distances undergo redshifts that render their arms blurry and hard to detect (Bamford et al. 2009). Land et al. and Bamford et al. were both able to identify these errors and correct for user biases by means of fairly simple algorithms.

With these measures in place, Galaxy Zoo's output exceeded all expectations. Comparing the project's classifications with those from three visual inspection studies conducted by professional astronomers on samples from the same SDSS images, Lintott et al. (2008) found that Zooites agreed with the experts in more than 90 % of cases—a rate comparable to experts' mutual agreement with one another. Similarly positive results have been reported for numerous crowdsourced projects across the natural sciences, including NASA's Clickworkers initiative (Kanefsky et al. 2001), Stardust@home (Méndez 2008), Foldit (Khatib et al. 2011b), and EyeWire (Kim et al. 2014).

It should perhaps come as no surprise to learn that large epistemic communities are capable of generating reliable observations for scientific research. After all, professors have long relied on untrained undergraduates for basic data collection tasks. The differences between that familiar case and this novel one are twofold. In the university setting, there are academic and social incentives to be a proficient data collector. In crowdsourced e-research, the data analysis platform itself ensures user performance. Second, we know by Condorcet's theorem that a jury's verdict asymptotically approaches truth as the number of better than random jurors increases. The quantity of participants involved in a given study therefore has a qualitative impact on the judgments they issue. The combination of shrewd web design and sheer user volume can turn the public into a valuable resource for scientific research.

### 3.2 Veritistic value and Bayesian reasoning

Goldman's social epistemology relies heavily on Bayesian inference, a methodology he argues is supported by the veritistic approach. He reports a result he credits to Shaked (see Goldman and Shaked 1991), who combines Bayes' theorem with Jensen's inequality to prove (roughly) that if agent $A$ has an accurate model of hypothesis $h$, then updating her beliefs with some relevant evidence $e$ will tend to bring $A$ closer to $h$'s truth-value. Specifically, he shows that, if the following three criteria are met:

(1) Relevance:

$$P(h) \neq P(h|e)$$

[2] The system worked as follows. For each galaxy $x$ to which volunteer $k$ assigned morphology $F$, the partial weight of $k$'s vote was defined as the number of other Zooites who agreed that $Fx$, divided by the total number of galaxies classified by $k$, $N_x(k)$. The summation of such ratios for all $N_x(k)$ represents $k$'s total weight $w_k$. Total weights for all users were then scaled to a mean of 1, and applied to each vote in the database. See Lintott et al. (2008).

(2) Bounds:

$$0 < C_A(h) < 1 \quad \text{and} \quad \frac{P(e|h)}{P(e|\sim h)} \neq 1$$

(3) Model accuracy:

$$C_A(h) = P(h) \quad \text{and} \quad \frac{C_A(e|h)}{C_A(e|\sim h)} = \frac{P(e|h)}{P(e|\sim h)}$$

then $A$'s expected change in veritistic value after conditionalising upon $e$ is strictly positive.[3] That is,

$$\mathrm{E}[V_A(h)|e - V_A(h)] > 0.$$

Shaked's theorem is rather trivial in most applications. Rarely do we have precise values of prior probabilities or relevant Bayes factors, and if we did, it would hardly be surprising to learn that combining the two would likely produce a net knowledge increase. Nevertheless, the result is important in the present context because, as we shall argue, it provides a firm logical foundation for crowdsourcing in the natural sciences.

Say $h$ stands for some particular observational claim, e.g. 'Galaxy $x$ is elliptical', and $e$ stands for a set of weighted user votes with respect to galaxy $x$'s morphology. When astrophysicist $A$ examines the data, she is in a good position to evaluate both the prior probability that $x$ is elliptical, given her background knowledge about the frequency of elliptical galaxies, and the likelihood ratio that $x$ is elliptical, given the degree of community consensus evident in $e$ and/or relevant user biases. Even if the quality of user contributions to some particularly confounding project were relatively low, as long as experts could determine their accuracy, then Shaked's theorem proves that Bayesian reasoning from such data will tend to increase the veritistic value of collective classifications. When amateur testimony is both accurately evaluated and generally reliable, as the protocols outlined above are designed to ensure, then the resultant data should be of extremely high quality.

[3] Our notation differs from that presented by Goldman and Shaked, but the substance of their theorem remains unchanged. See Goldman and Shaked (1991). Their complete proof only appears in the appendix of a later book, which includes a reprinted edition of Goldman and Shaked's original article. See Goldman (1992, chapter 12).

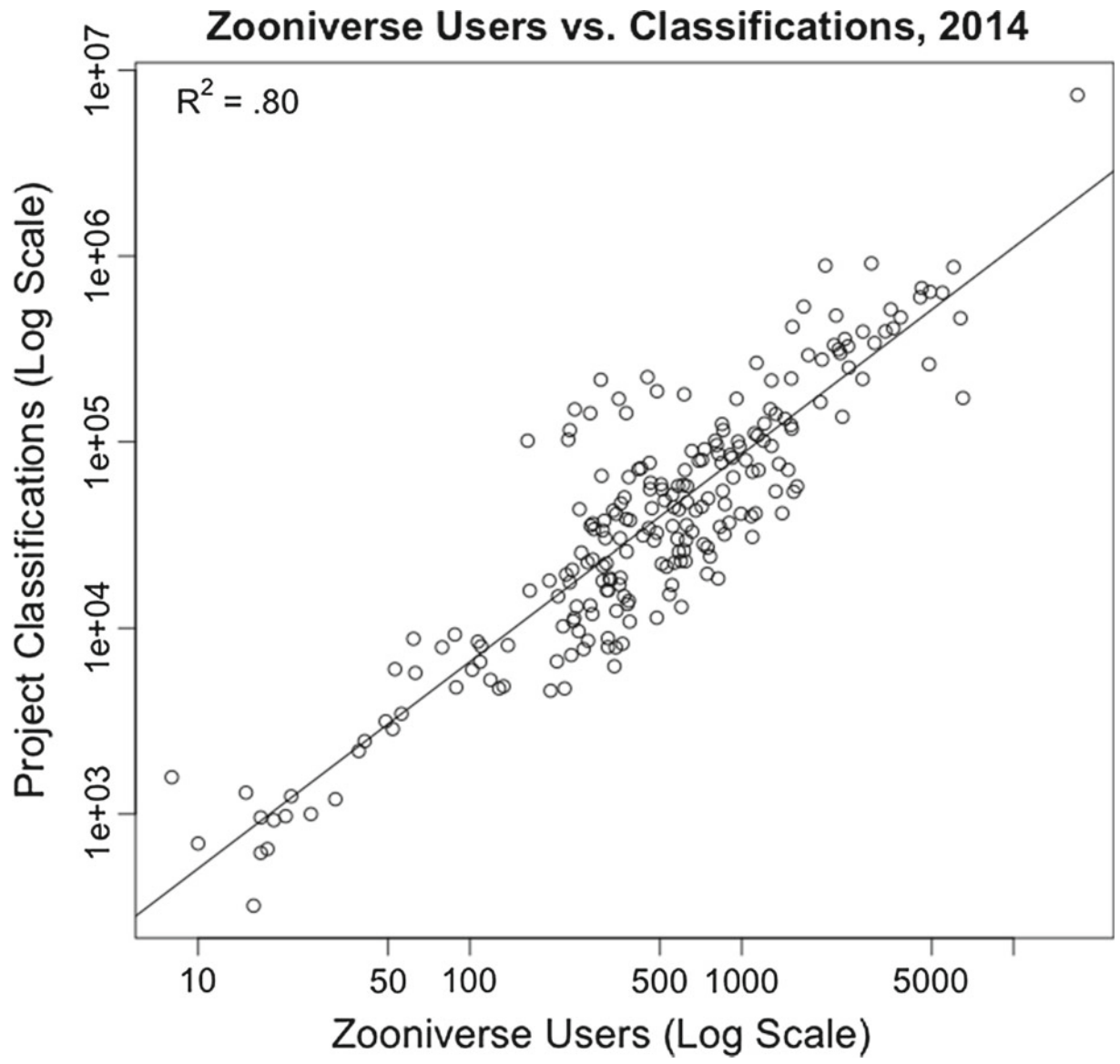

**Fig. 3** Log–log scatterplot of Zooniverse users versus classifications, with an ordinary least squares regression line fit to the data

## 4 Scalability: the more the Merrier

High-throughput techniques across the natural sciences have given modern researchers more accurate, precise, and numerous measurements than ever before, yet pattern recognition software for visual, audio, and video data is still fairly crude. How can scientists take advantage of these emerging technologies most efficiently?

### 4.1 Users and observations in Zooniverse

In the months preceding the launch of Galaxy Zoo, Zooniverse cofounder Kevin Schawinski spent a week classifying 50,000 galaxies as part of his D.Phil. research in the Astrophysics Department at the University of Oxford (Schawinski et al. 2007). The task was gruelling. Presuming 12-hour workdays, Schawinski must have averaged a classification every six seconds for seven straight days. By comparison, the day Galaxy Zoo went online, users were averaging 70,000 classifications per hour (Nielsen 2011). By the time Zooites finished processing the complete SDSS survey of almost 900,000 objects, their work constituted the largest morphological catalogue in the history of astronomy (Bamford et al. 2009).

Zooniverse's 2014 transaction logs reveal a strong positive correlation between a project's user totals and the number of classifications it generates. Figure 3 is a log–log scatterplot depicting the relationship between these two variables over the 223 complete project-months for which such data were recorded. A simple linear regression model was fit to the log transform data, indicating that user totals account for approximately 80 % of the variance in a project's classificatory output. While variables like user engagement and media coverage would no doubt help to construct a more complete picture of how and why some citizen science initiatives are more fruitful than others (Cox et al. 2015), this plot clearly shows that the number of volunteers who contribute to a project is a strong predictor of how many observations it will produce.

The success of any given citizen science project has always been dependent on its ability to attract sufficient volunteers. However, only in the era of global ICT networks can these initiatives reach the critical mass at which they begin to match or even surpass the efforts of professionals relying on more traditional modes of data processing. Consider, for example, the case of astronomical catalogues. An astronomical catalogue is a complete list of objects of some common type (e.g., galaxies) detected by one or several instruments working in concert, usually as part of an astronomical survey (e.g., the SDSS). While scientific articles often draw on select or simulated data to explore some particular phenomenon, astronomical catalogues represent researchers' total observational output of a particular kind. Comparing the number of observations in traditional and crowdsourced editions of such works therefore offers the best means of testing the relative fruitfulness of the two methodologies.

In the four years since the aforementioned Galaxy Zoo catalogue was published, Zooniverse has gathered user classifications into seven more astronomical catalogues, two of which were the first of their kind.[4] The other five are the largest of their sort ever compiled, exceeding previous record holders by more than order of magnitude on average. By comparison, traditional catalogues tend to build on previous work in increments of about 80 %. Table 1 includes observation totals for each of these five Zooniverse catalogues[5] and the traditional catalogues they superseded,[6] along with the percent increases in observation counts represented by each. Where possible, statistics on three previous catalogues are included for comparison.

A Kolmogorov–Smirnov (K–S) test found significant difference between the percent increases in observation totals represented by Zooniverse projects and those of traditional catalogues relative to previous collections, $D = 0.86$, $p = 0.02$. While the sample size in this analysis is admittedly small, the effect size detected is very large, Cohen's $d = 1.22$, demonstrating a difference of more than a full standard deviation between the two groups' means. Given the strength and uniformity of these results, we may confidently conclude that crowdsourcing is categorically superior to traditional visual inspection methods at gathering large quantities of empirical evidence for astronomical studies. Similar results have been reported for large-scale ecology projects (Swanson et al. 2015).

[4] Following the discovery of a rare object in the initial Galaxy Zoo project (about which more below), Zooniverse launched an intergalactic search for similar anomalies, ultimately resulting in the identification of 19 candidate 'voorwerps' (Keel et al. 2012). Though there are several other coronal mass ejection (CME) catalogues, Zooniverse's is unique in that it deliberately prioritises quality over quantity, ignoring minor CMEs while gathering the most extensive time series data ever recorded on a relatively small number of notable solar events (Barnard et al. 2014). Note that, because neither Zooniverse project bears quantitative comparison with any traditional catalogue, both are excluded from the following analysis.

[5] Galaxy Zoo 1 gathered basic galactic morphologies (Lintott et al. 2011); Galaxy Zoo 2 was devoted to detailed galactic morphologies (Willett et al. 2013); results from both projects were used to create a catalogue of overlapping galaxies (Keel et al. 2013); the Milky Way Project found infrared bubbles in our own galaxy (Simpson et al. 2012a); and the Andromeda Project sought stellar clusters in our neighbouring Andromeda galaxy (Johnson et al. 2015).

[6] All previous observations of overlapping galaxies are catalogued in Appendix A of (Keel et al. 2013); traditional catalogues of infrared bubbles were compiled by Churchwell et al. (2006, 2007); the three largest collections of basic galactic morphologies gathered by traditional means are all due to Schawinski et al. (2007); Fukugita et al. (2007), Baillard et al. (2011), and Nair and Abraham (2010) used visual inspection to catalogue detailed galactic morphologies of increasing size; and the three largest stellar cluster catalogues compiled before Zooniverse were published by Bastian et al. (2012), San Roman et al. (2010), and Popescu et al. (2012), respectively.

**Table 1** Observation totals and percent increases across five different types of astronomical catalogues

| Catalogue | Method | Observations | % Increase |
|---|---|---|---|
| Overlapping Galaxies | Traditional | 25 | |
| | Crowdsourcing | 1990 | 7860 |
| Infrared Bubbles | Traditional | 322 | |
| | Traditional | 591 | 83.54 |
| | Crowdsourcing | 5106 | 763.96 |
| Basic Galactic | Traditional | 15,729 | |
| Morphologies | Traditional | 19, 649 | 24.92 |
| | Traditional | 48,023 | 144.40 |
| | Crowdsourcing | 738,175 | 1437.13 |
| Detailed Galactic | Traditional | 2253 | |
| Morphologies | Traditional | 4458 | 97.87 |
| | Traditional | 14,034 | 214.80 |
| | Crowdsourcing | 304,122 | 2067.04 |
| Stellar Clusters | Traditional | 751 | |
| | Traditional | 803 | 6.92 |
| | Traditional | 920 | 14.57 |
| | Crowdsourcing | 2753 | 199.24 |

### 4.2 Epistemic communities and the principle of total evidence

The plot in Fig. 3 suggests that observations are a monotonically increasing function of users in Zooniverse. Note that the deliberate redundancy mentioned in Sect. 3.1, whereby each datum is classified numerous times by various users, has no bearing on the regression line's slope or residual error. The only parameter subject to change, should all values of the dependent variable be divided by some constant (say, 38), would be the line's intercept, as the data points would all shift downward with no impact on the model's goodness of fit. This direct relationship between a project's contributors and its data processing power is strong evidence in favour of crowdsourcing's scientific utility. As we have seen, the largest astronomical catalogues ever collected were made with the assistance of hundreds of thousands of volunteers.

The value in maximising relevant data for empirical analyses is widely recognised, though rarely does the practice receive explicit justification. Bernoulli (1713) was perhaps the first to write that probability calculations require the use of all available evidence. Keynes built upon this view, arguing that, while new observations may raise or lower the likelihood of a given hypothesis, they invariably increase what he called ‘the weight of evidence’, leading to ‘more substantial’ conclusions (1921, p. 77).[7] Carnap upgrades this proposal to a full blown principle, claiming that ‘In the application of inductive logic to a given knowledge situation, the total evidence available must be taken as basis for determining the degree of confirmation’ (1950, p. 221). Though some have challenged Carnap on this point (Ayer 1957; McLaughlin 1970), the vast majority of philosophers, statisticians, and laypeople alike tend to view the principle of total evidence (TE) as little more than common sense (Hempel 1960; Efron 2010).

There are several compelling reasons to accept TE. Increased sample sizes improve the accuracy and precision of statistical estimates and inferences, narrowing the confidence intervals around predictions and parameters, thereby limiting the likelihood of Type I and Type II errors. The epistemological merits of TE can be formalised in a Bayesian framework using Shaked’s theorem. Let $e$ stand for some collection of observations, say of galactic morphologies. Let $e^*$ stand for some larger body of similar observations, say twice as many galactic morphologies. Let $h$ stand for some relevant hypothesis, perhaps pertaining to the distribution of galactic morphologies. Then while $e^*$’s superior weight alone does not entail any conclusions regarding the relative values of the conditional probabilities $P(h|e)$ and $P(h|e^*)$, we can be more confident in the latter evaluation than in the former. It follows from Shaked’s theorem that heavier bodies of evidence will tend to increase the veritistic value of our judgment in $h$. Provided the following modified conditions are met:

(1) Relevance:

$$P(h) \neq P(h|e) \text{ and } P(h) \neq P(h|e^*)$$

(2) Bounds:

$$0 < C_A(h) < 1, \ \frac{P(e|h)}{P(e|\sim h)} \neq 1, \ \text{ and } \ \frac{P(e*|h)}{P(e*|\sim h)} \neq 1$$

(3) Model accuracy:

$$C_A(h) = P(h), \ \frac{C_A(e|h)}{C_A(e|\sim h)} = \frac{P(e|h)}{P(e|\sim h)}, \ \text{ and } \ \frac{C_A(e*|h)}{C_A(e*|\sim h)} = \frac{P(e*|h)}{P(e*|\sim h)}$$

then what holds for prior and posterior probabilities in Shaked’s theorem will hold for beliefs updated with $e$ and $e^*$, respectively. That is, we may derive the following inequality:

[7] The term ‘weight of evidence’ is employed in a very different sense by Good (1983), and still another by Joyce (2005). In what follows, we adopt the Keynesian terminology. See Joyce (2005) for an insightful breakdown of the subtle distinctions between various interpretations of evidentiary weight, balance, and specificity in Bayesian contexts.

$$\mathrm{E}[V_A(h)|e^* - V_A(h)|e] > 0.$$

The intuitive appeal of TE now becomes clear. An epistemic agent conditionalising upon a relatively large collection of observations is more likely to be right about a relevant hypothesis than she would be given a smaller body of similar evidence. This result goes hand in hand with Good's theorem (1967), which purports to prove that rational agents must maximise free evidence, although his argument relies upon extra premises that we do not consider here.

Gathering as many observations as possible for scientific investigation is not just a matter of fine-tuning particular models. Large samples are more likely to contain anomalous data, which numerous historians and philosophers of science point out are crucial for theoretical progress. Such unexpected discoveries may falsify prevailing hypotheses (Popper 1959) or perhaps even help inaugurate a new research paradigm (Kuhn 1962). Since anomalous observations are, by definition, low probability events, we should only expect to find them in large datasets. While one or two anomalies could plausibly be dismissed as mere outliers, the accumulation of rare data in large sample sizes makes their presence more salient and their need for explanation more pressing.

Given the results of the regression in Sect. 4.1 and the preceding defence of TE, it is tempting to conclude that veritistic value is a monotonically increasing function of epistemic community size. Yet the generality of this claim is constrained by two factors: the nature of a particular scientific investigation, and the technology available to those who undertake it. The Zooniverse model is only applicable to projects with intractable amounts of data that require little or no expertise to process. This describes a large and diverse but by no means exhaustive set of scientific studies. Virtual citizen science also presumes a technological context in which computational resources are sufficiently advanced to establish a global crowdsourcing platform, but cannot (yet) be used to reliably automate the tasks put forward to volunteers. Numerous groups, including members of the Zooniverse team, are hard at work to create software that will render the user classification system obsolete (Banerji et al. 2010; Simpson et al. 2012b; Shamir et al. 2014). Zookeepers predict that, even once such programs are employed, volunteers will remain a valuable part of e-research, helping to refine algorithms through anomaly detection and review (Clery 2011; Fortson et al. 2012).

When it comes to participation in citizen science, the more the merrier. Only online platforms offer the kind of scalability required to host hundreds of thousands of volunteers for any given project, and only at these volumes does the data processing power of untrained amateurs begin to compete with (or exceed) that of experts using traditional observation methods. The combination of high quality and high quantity data is essential for scientific confirmation and discovery.

## 5 Connectivity: *E Pluribus Unum*

The reliability and scalability of crowdsourced e-research has helped amass enormous volumes of reliable observations across the natural sciences. But are the methodology's contributions limited to clever web design and evidentiary archiving, or does

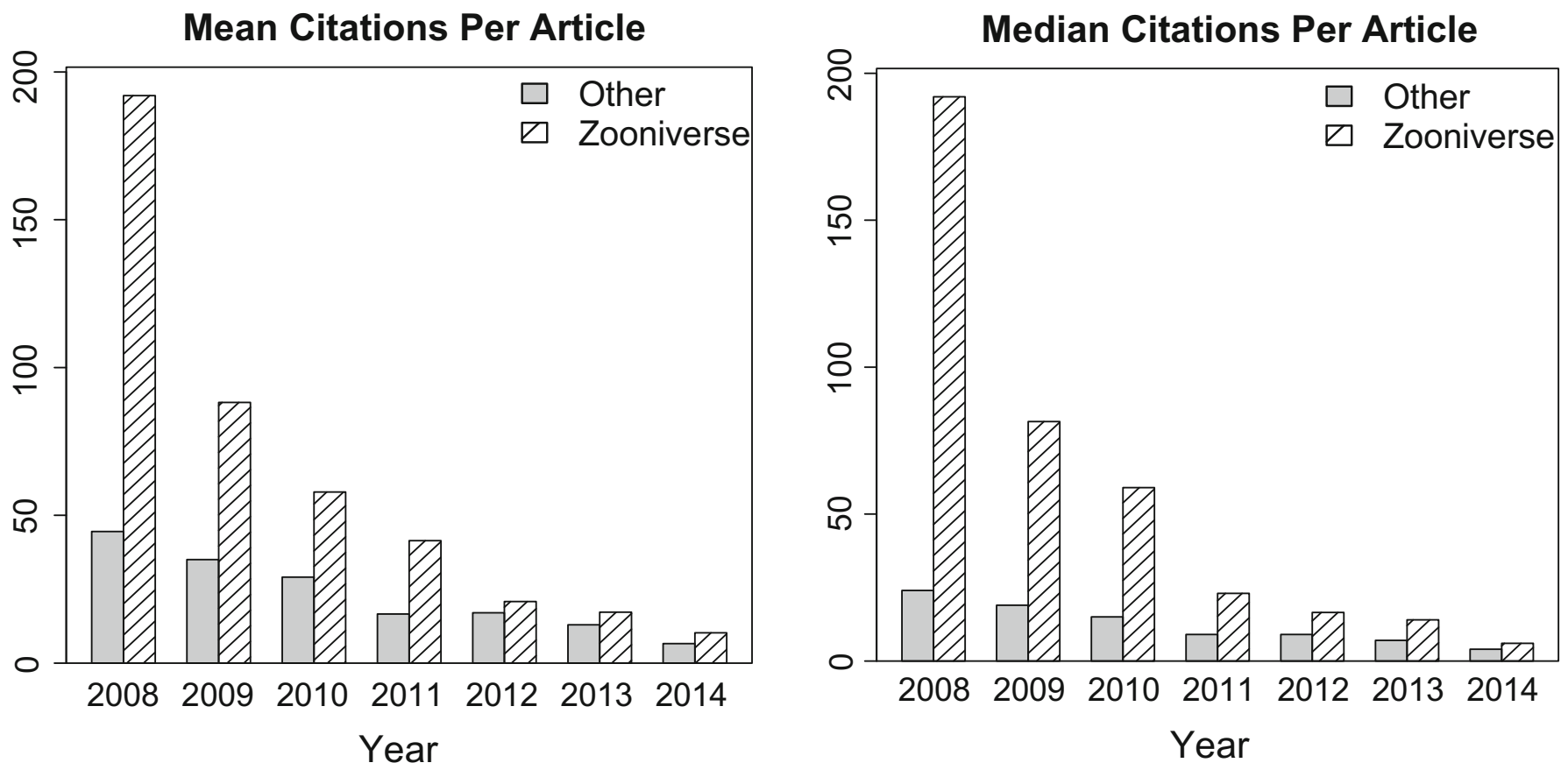

**Fig. 4** Bar plots comparing mean and median citations per article for Zooniverse and other sources using the same raw data. Since academic citations are usually power-law distributed (Barabási 2002), the median is probably a more reliable measure of central tendency than the mean for these distributions

crowdsourcing hold promise for more substantial forms of scientific knowledge as well?

### 5.1 Scientometric performance

The quality of a scientific discovery is notoriously difficult to quantify. However, the analytic tools of scientometrics provide several methods for attempting to do so (Price 1963; Leydesdorff 2001). Because the majority of Zooniverse projects draw their raw data from public access archives, such as the SDSS and the Hubble Space Telescope, other papers by scientists using the same source materials constitute the most natural control group for scientometric analysis. Of the 68 Zooniverse articles published before 2015, 62 were the result of projects that relied exclusively on publicly available data. In the same timeframe, other scientists published 5522 articles using the same sources.

Comparing the citation and journal data of these two groups provides some insight into the relative influence of Zooniverse's scientific output.[8] A simple technique of weighing the two samples against each other is through the common scientometric indicator of citations per article. This statistic is biased towards older articles for obvious reasons, which accounts for the steep drop off over time evident in Fig. 4. However, both charts reveal another clear trend. Without exception, Zooniverse's papers are consistently more cited on average than those by scientists using traditional research methodologies to investigate the same material. While the large discrepancy in 2008

[8] Because Zooniverse has been widely studied by sociologists, only citations from natural science journals were counted for this comparison. The true influence of Zooniverse publications in fact extends beyond this narrowly circumscribed academic domain.

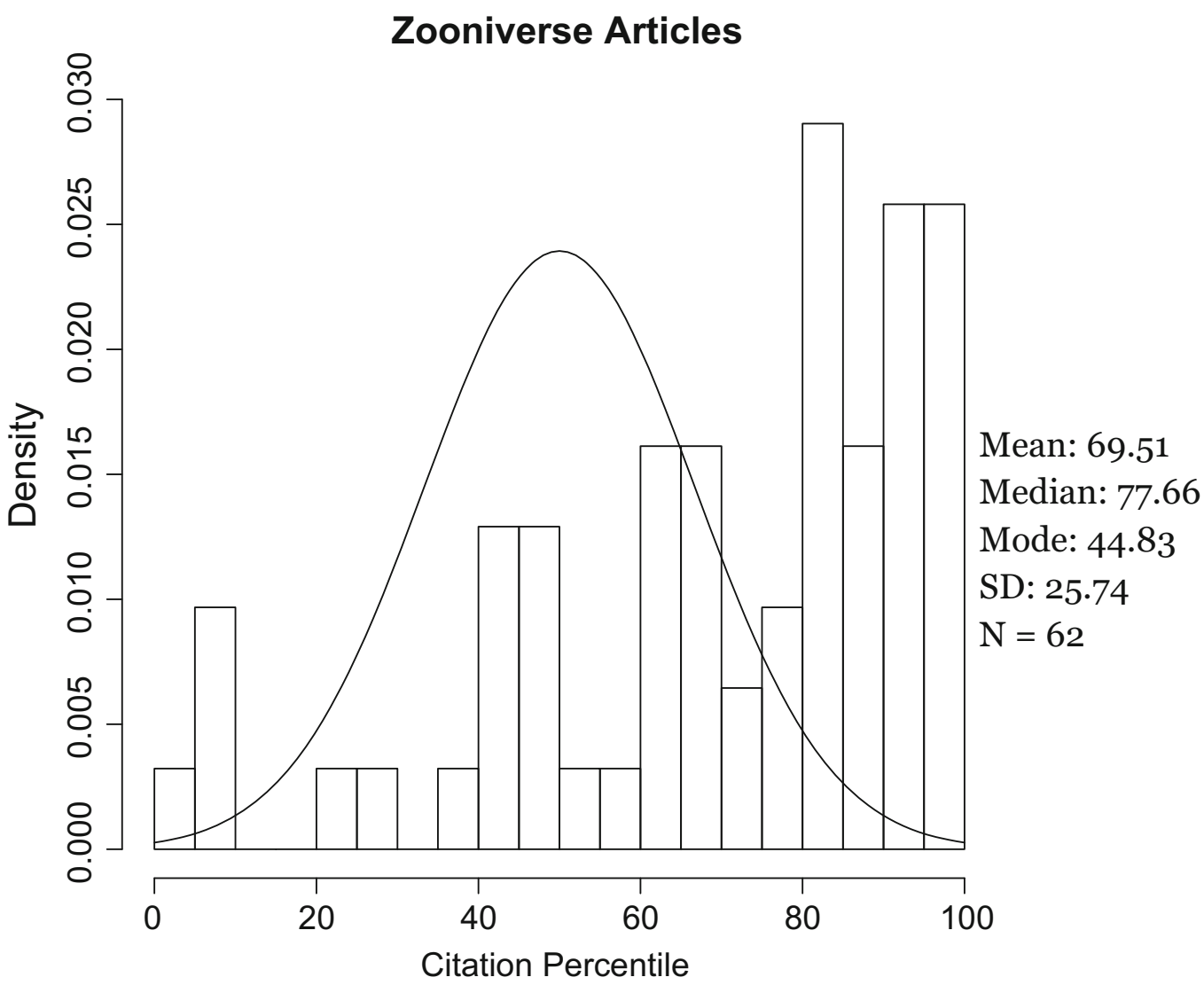

**Fig. 5** Histogram depicting the distribution of citation percentiles across all Zooniverse articles published from 2008 to 2014. A normal curve $N(50, 16.67^2)$ is overlaid for comparison, with parameters chosen so as to centre the distribution at the middle of the citation percentile range and let all points under the curve on [0, 100) fall within three standard deviations of the mean

is likely due to the substantial buzz around the first Galaxy Zoo article, these bar plots demonstrate that the trend has remained remarkably persistent over time.

We might expect that the citation percentiles by year and data source for a theoretical 'average' lab would tend to follow an approximately normal distribution, with a small but roughly equal number of articles performing very well and very poorly, and the vast majority falling somewhere in between. If so, then we can confidently assert that Zooniverse is not an average lab. Fig. 5 is a histogram of Zooniverse's citation percentiles, with a normal curve overlaid for comparison. We find here that nearly half of all Zooniverse papers are in the top quintile of most cited articles for their year and data source, with more than a quarter in the top 10 %. A K–S test found significant deviation between these observed results and those expected of a normal distribution, $D = 0.48$, $p < 0.001$.

The distribution of Zooniverse's citation percentiles has a skewness of $\gamma_1 = -1.04$, reflecting a high incidence of papers in the upper ranges of most cited articles for their year and data source. By contrast, the distribution of citation percentiles for the 5522 articles in the control group is nearly uniform. The dissimilar shapes of the two distributions are clearly visualised in Fig. 6, where density plots for both are overlaid for comparison. We find here that articles by researchers using traditional methodologies are more concentrated below approximately the 50th percentile, while Zooniverse papers are more likely to be found in the upper half of the data range. A K–S test found significant difference between the two groups, $D = 0.35$, $p < 0.001$.

Zooniverse's influence is rivalled only by that of the most prestigious labs in the field. Of the 5522 articles in the control group, 136 were published by researchers affil-iated with the University of Cambridge, home to one of the most esteemed astronomy institutes in the world. The distribution of citation percentiles for these papers is negatively skewed, $\gamma_1 = -0.43$, as one might expect—but less so than that of Zooniverse articles, indicating that the latter are more likely to have higher citation percentiles than the former. A K–S test on the two distributions found no statistically significant difference between them, $D = 0.18$, $p = 0.12$, suggesting that Zooniverse's citation percentiles could plausibly represent a random sampling of Cambridge's.

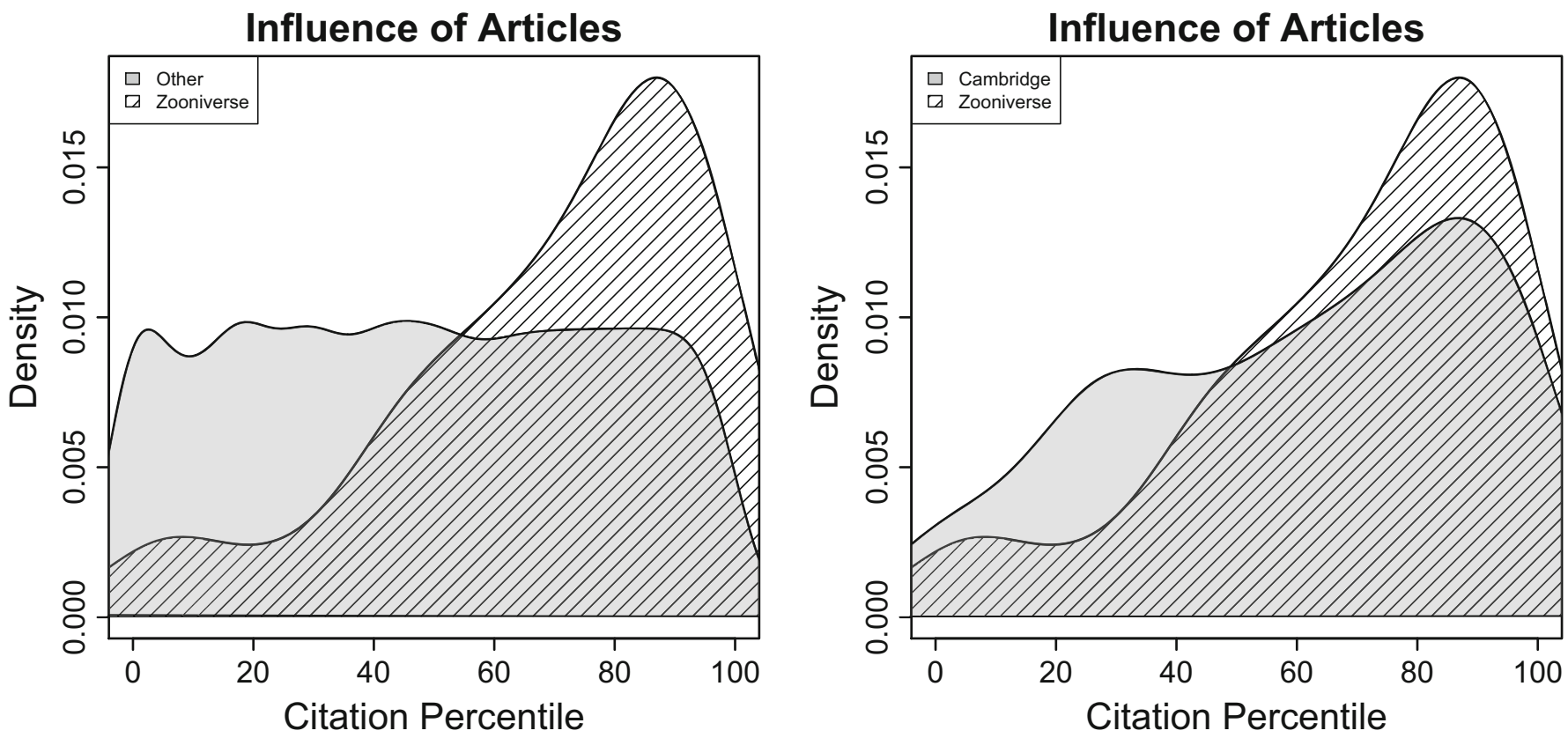

**Fig. 6** Density plots representing the distribution of citation percentiles for Zooniverse articles versus those by all others using the same raw data, and Cambridge researchers using the same raw data, respectively

The disparity in article influence between Zooniverse's publications and those from the general population cannot be accounted for by journal data alone. Systematic comparison of the average impact factor[9] and h-index[10] of both groups' top ten most frequent publishers of articles weighted by output for each year between 2008 and 2014—journals that cumulatively account for over 90 % of all such material—demonstrates that Zooniverse had no systematic advantage in academic visibility to bolster its citation numbers.

While the two statistics visualised in Fig. 7 do not perfectly coincide, they both reflect a broadly similar state of affairs. By either measure, Zooniverse's publishers are roughly as influential as those of other researchers using the same data sources over time. K–S tests on the two pairs of weighted averages found insignificant differences between the distributions, with Zooniverse's journals tending to have marginally lower impact factors, $D = 0.57$, $p = 0.21$, and h-indexes, $D = 0.43$, $p = 0.54$, on average.

Of course, the true value of a scientific discovery is impossible to measure. It corresponds to an abstract and subjective concept that evolves over time and has no clear operationalisation. However, it is hard to imagine how Zooniverse publications

9 A journal's impact factor refers to the ratio of its total number of articles cited by other indexed publications within the past two years, and the total number of articles published by that journal in the past two years (Garfield 1972). Impact factor data for 2008–2014 was gathered from the ISI Journal Citation Reports.

10 A journal's h-index is defined as its number of articles $h$ that have each been cited in other journals at least $h$ times (Hirsch 2005). H-index data for 2008–2014 was compiled from Elsevier's Scopus database.

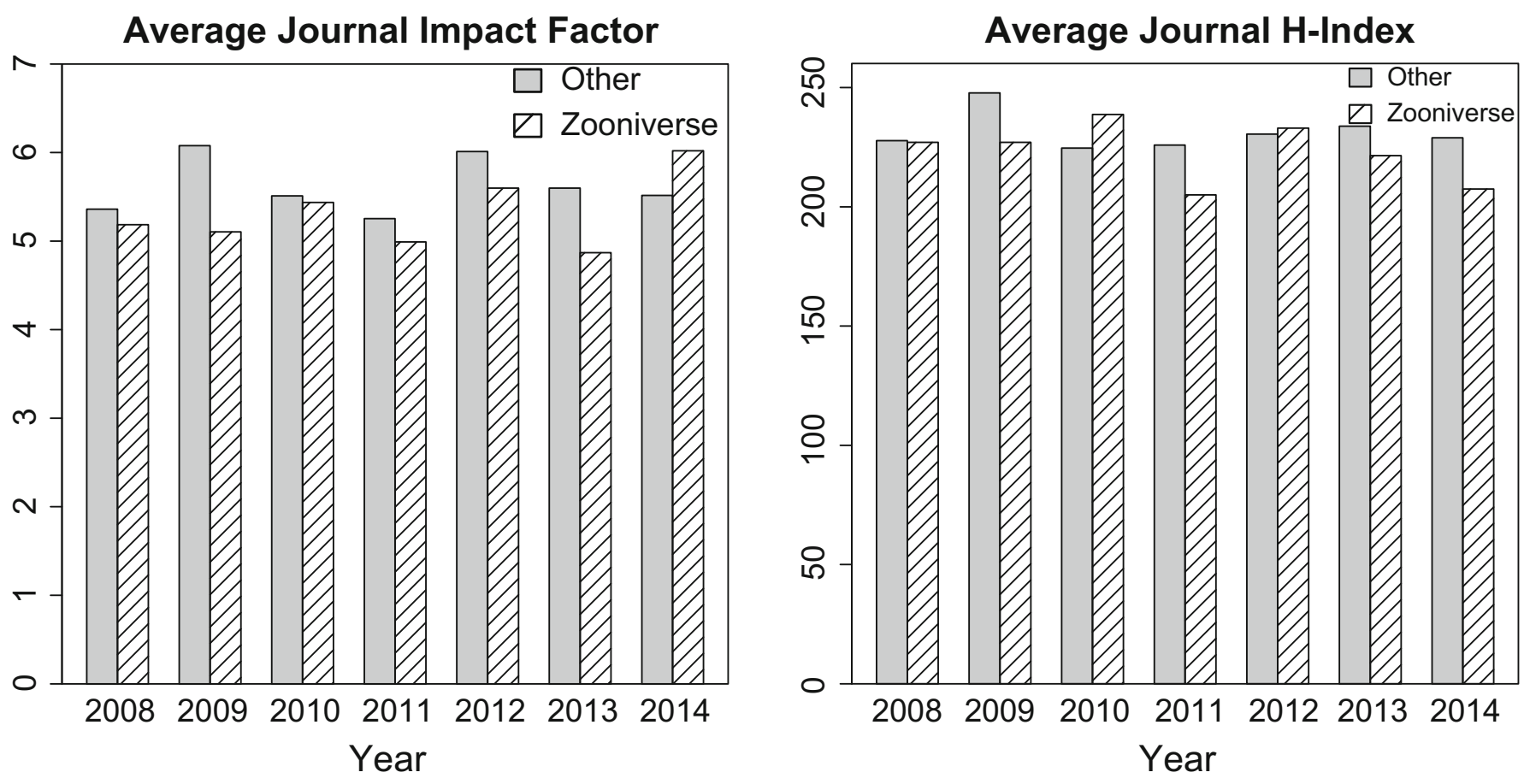

**Fig. 7** Bar plots comparing the mean impact factor and h-index values for publishers of Zooniverse papers with those by others using the same raw data

could so consistently outperform those by other labs in the same field using the same data if they did not at least sometimes contain substantive contributions to scientific discourse. This minimal claim is all that is required to answer our question at the top of Sect. 5 in the affirmative. Crowdsourcing can and does produce high quality science beyond mere data aggregation.

### 5.2 Network architecture and distributed knowledge

The quality and quantity of observations gathered by Zooniverse no doubt factors into the strong scientometric performance of their publications over time. Novelty and good publicity may also play a role (Cox et al. 2015). But it is the structure of the site's sociotechnical network that truly enables principal investigators to harness the community's resources for maximal discovery value. Some of Zooniverse's most important contributions have been the result of confused users taking to the site's talk forums to discuss strange objects that did not seem to fit into any of the available categories for classification. That was the case with 'Hanny's voorwerp', a large cloud of bright green gas in the constellation Leo Minor, which researchers believe may be the first quasar light echo ever observed (Lintott et al. 2009). User comments also led to the discovery of so-called 'green pea galaxies' (Cardamone et al. 2009), triple mergers (Darg et al. 2011), supernovas (Smith et al. 2011), and overlapping galaxies (Keel et al. 2013) in SDSS data. Reviewing the results of their inaugural project, the site's founders concluded that 'The Galaxy Zoo forum has been a scientific gold mine' (Fortson et al. 2012, p. 226).

Zooites not only classify the objects provided by Zookeepers, but flag anomalies for further discussion. The intermingling of diverse views and levels of expertise in the Zooniverse talk forums naturally drives expert attention toward the most deserving data (Page 2007). In several cases, researchers have used those findings to launch new

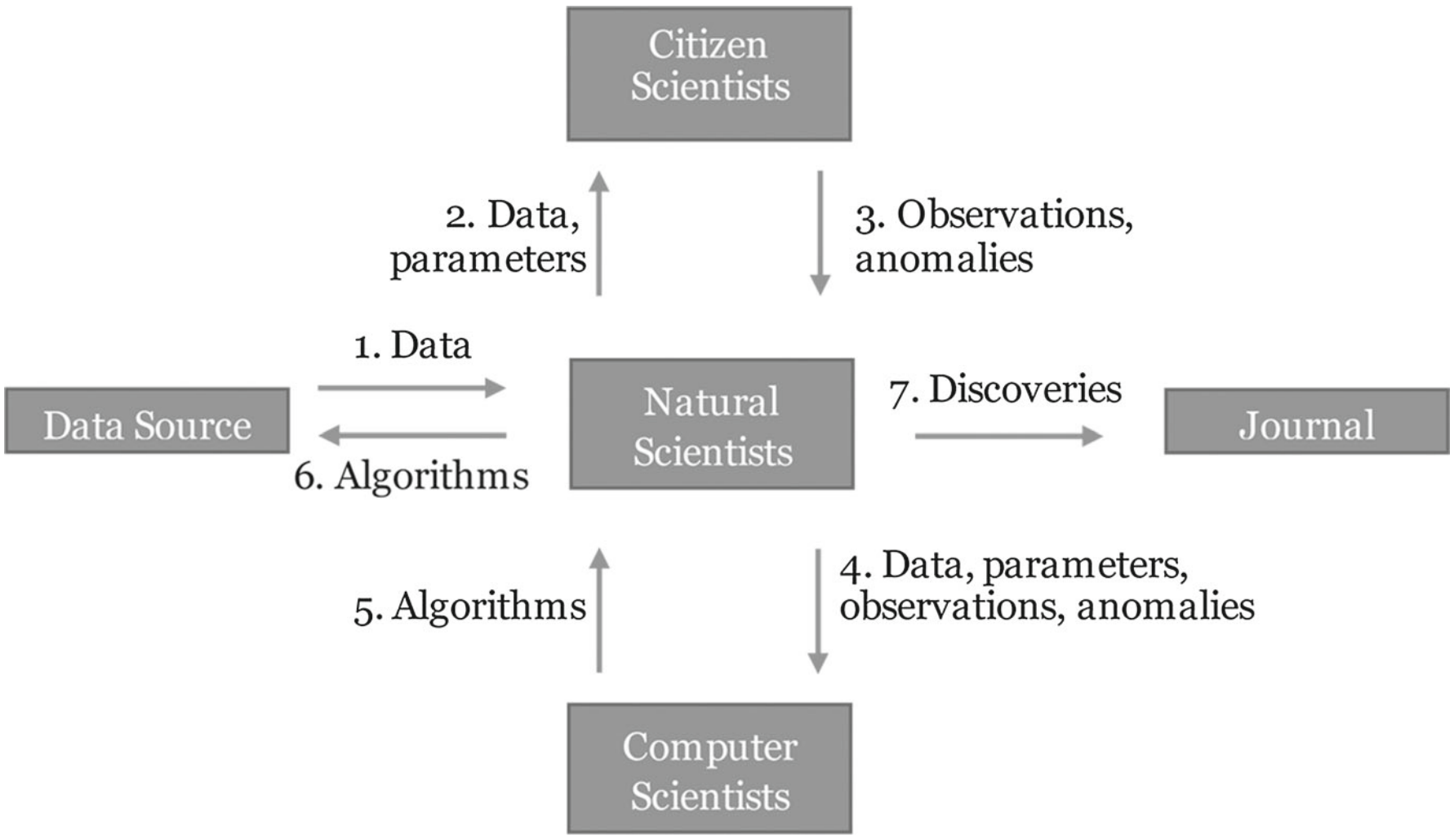

**Fig. 8** Diagram of sociotechnical knowledge production in Zooniverse. The first four nodes of the network (i.e., every step prior to sending discoveries to a journal) form a recursive loop that results in increasingly refined observational results

projects that branch off from earlier ones in pursuit of similar rare objects. This process demonstrates the new and remarkable ways in which amateurs, experts, and digital technologies come together to form a cohesive sociotechnical system in crowdsourced science projects. Figure 8 depicts the knowledge production network in Zooniverse. Note how computer-mediated human cognition at the nodes is transferred by ICTs at the edges, creating a complex epistemic system that refines and curates observations until ready for publication. Such recursive patterns of discovery are indicative of a mature and fruitful scientific methodology.

The sociotechnical network depicted above is designed to unlock the distributed knowledge of Zooites and Zookeepers. The formal definition of distributed knowledge was originally proposed by Halpern and Moses (1990) and later refined by Fagin et al. (1995). A complete explication of the semantics for their model of epistemic logic is beyond the scope of this paper, but the basic idea is fairly intuitive. Their distributed knowledge operator $D$ is defined in such a way that, for some group of agents $G$, $D_G$ represents not only the sum of all things known by $G$'s members, but also all valid entailments of their pooled knowledge.[11]

A version of Fagin et al.'s logic is widely implemented in multiagent computing (Wooldridge 2002), and has clear applications to any form of collaborative research. For instance, it helps defuse the philosophical puzzles that arise when large teams of experts produce results that no single one of them fully understands. Hardwig (1985) calls this *the problem of epistemic dependence*, and proposes an elaborate theory of justification in his effort to salvage the primacy of individual knowledge. Longino

[11] Say Alice knows that either 3 or 4 is prime. Bob is unsure about 3, but he is certain that 4 is not prime. Then even though neither Alice nor Bob alone knows that 3 is prime, together they could deduce this fact. The knowledge that 3 is prime is *distributed* between Alice and Bob, whether they realise it or not.

(1990, 2001) challenges Hardwig's epistemic individualism, arguing that cognitive processes are essentially social, and therefore that individual knowledge itself is either emergent or misconstrued.

Neither alternative is particularly compelling. Longino's account has counterintuitive consequences for philosophy of mind, while Hardwig's appears to be based on a metaphysical misunderstanding. His reticence to grant epistemic agency to an entire research group is probably rooted in the metatheoretical desire for ontological parsimony. If we have already acknowledged the existence of agents $A$ and $B$, then we would rather avoid countenancing the existence of some third agent $C$ such that $C = A \cup B$. It is not entirely clear, however, what metaphysical commitments accompany propositions like 'The jury finds the defendant guilty', 'The army won the battle', 'The class is on a field trip', etc. Multiagent systems are regularly treated as perfectly ordinary epistemic (Goldman 2003) and indeed moral subjects (Floridi 2013).

Mereological subtleties and confusions abound in the natural sciences, not least because it is often difficult or impossible to establish the ideal unit of analysis (Winther 2011). The question of when to assign collective agency to a group of individuals raises particularly vexing issues in biology (Jones 2017), not to mention moral philosophy (Searle 1990). Some notable philosophers argue that all talk of aggregation is essentially pragmatic, with little or no ontological implications. For example, Hume (1748/2008) writes that 'the uniting of…parts into a whole, like the uniting of several distinct countries into one kingdom, or several distinct members into one body, is performed merely by an arbitrary act of the mind, and has no influence on the nature of things' (9.11/65). Wittgenstein echoes this sentiment, rejecting the notion that there are any objective distinctions to be drawn between parts and wholes. 'To the *philosophical* question: "Is the visual image of this tree composite, and what are its component parts?" the correct answer is: "That depends on what you understand by 'composite'." (And that is of course not an answer but a rejection of the question)' (1953, § 47).

The sociotechnical network may not be a metaphysical entity per se, but its epistemic agency is explanatorily essential to the knowledge it generates at the system level of abstraction (Floridi 2011; Winther 2011). The mere aggregation of Zooniverse's units—some users here, a mainframe there—does not begin to account for the site's consistent output of high impact scientific publications. It is the complete sociotechnical process, not a summation of localised knowers, that leads to new and influential discoveries in crowdsourced e-research. Proper coordination is essential (Floridi 2004). Cautious philosophers who accept the notion of distributed cognition but balk at the idea of extended or collective agency (e.g., Giere 2007) are insisting on a distinction without a difference. Drawing circles around every individual involved in these projects and declaring that agency can only exist within those borders is as arbitrary as it is unnecessary (Longino 2013). Epistemic agency supervenes upon the people and technology of which the sociotechnical system is comprised, leveraging both human intelligence and computational resources. Crowdsourcing is hardly the only activity in which this kind of heterogeneous connectivity is evident (Hutchins 1995; Cetina 1999; Latour 2005), but it does pose a vivid example of how large groups come together to forge scientific knowledge.

Crowdsourced science may constitute a radical departure from traditional research methodologies, but its most interesting features lie not in what it *adds* to scientific inquiry so much as what it *reveals* about it. Note how technology permeates every step in the knowledge production chain diagrammed in Fig. 8. Not only do the arrows depict the flow of information through ICT networks, but at every node people use computers to generate, analyse, simulate, and/or disseminate information to other nodes. While epistemologists over the last few decades have begun to focus on the social aspects of science, comparatively little attention has been paid to its technological underpinnings.

The very act of measurement itself, perhaps the most fundamental of all scientific activities, requires at least some minimal tools. Especially in the natural sciences, where sophisticated instruments are increasingly operated by computers, simulation has become an essential research methodology, and large groups of collaborators frequently share data via online networks, there can be no denying that technology functions as a mediating, even constitutive component of epistemic systems.

## 6 Conclusion

Statistical analysis of Zooniverse's publications and user activity indicates that crowdsourcing is a uniquely reliable, scalable, and connective method of generating scientific knowledge. This empirical evidence is supported by Bayesian reasoning within an epistemological framework that seeks to maximise the expected veritistic value of scientific hypotheses. Our work clarifies the philosophical foundations of virtual citizen science and highlights the irreducibly sociotechnical component of scientific research.

Collaboration and computation are ubiquitous across the natural sciences, and have been for decades. The recent popularity of websites like Zooniverse is a salient reminder of how potent the combination of large epistemic communities and well-designed technologies can be. The philosophical implications of this union have not gone completely unremarked (see Cetina 1999; Clark 2008; Floridi 2011), and some recent unpublished doctoral dissertations (e.g., Zollman 2007; Simon 2010) suggest that it may be a growing area of research. Further investigation of science's sociotechnical nature will prove fruitful for theorists and practitioners alike.

We cannot be certain just what scientific developments the future holds in store, but we can be confident that many of our next great discoveries will be made thanks to some complex partnership of minds and machines. Whether or not such results are the product of crowdsourcing, thorough investigation of this strange and remarkable methodology sheds new light on the varied modes of human knowledge. Clearly the time has come to endorse a sociotechnical turn in the philosophy of science that combines insights from statistics and logic to analyse the latest developments in scientific research.

**Acknowledgments** The authors would like to thank David Kinney for his insightful comments on earlier drafts of this article. We also thank our anonymous referees for their numerous helpful recommendations.